# Observations on a series of merging magnetic flux ropes within an interplanetary coronal mass ejection

Hengqiang Feng[1*], Yan Zhao[1], Guoqing Zhao[1], Qiang Liu[1], Dejin Wu[2]

[1]*Institute of space physics, Luoyang Normal University, Luoyang, China*

[2]*Key Laboratory of Planetary Sciences, Purple Mountain Observatory, CAS, Nanjing, China,*

[*]*fenghq9921@163.com*

**Abstract:** Coronal mass ejections (CMEs) are intense solar explosive eruptions. CMEs are highly important players in solar-terrestrial relationships, and they have important consequences for major geomagnetic storms and energetic particle events. It has been unclear how CMEs evolve when they propagate in the heliosphere. Here we report an interplanetary coronal mass ejection (ICME) consisting of multiple magnetic flux ropes measured by WIND on March 25–26, 1998. These magnetic flux ropes were merging with each other. The observations indicate that internal interactions (reconnections) within multi-flux-rope CME can coalesce into large-scale ropes, which may improve our understanding of the interplanetary evolution of CMEs. In addition, we speculated that the reported rope-rope interactions may also exist between successive rope-like CMEs and are important for the space weather forecasting.

## 1. Introduction

Coronal mass ejections (CMEs) are intense solar explosive eruptions that eject large amounts of plasma and magnetic field from the solar atmosphere, and are known to be the main source of intense geoeffectiveness (*Brueckner et al.*, 1998; *Tsurutani et al.*, 1988). CMEs are usually assumed to have magnetic flux rope structures near the Sun because of their helical shapes (e.g., *Rust & Kumar* 1996; *Canfield et al.*, 1999; *Liu et al.*, 2010; *Zhang et al.*, 2012). However, observations at 1 AU reveal that only 30%–40% of interplanetary CMEs (ICMEs) have the appearance of ropes and the ratio strongly depends on the solar activity (*Bothmer and Schwenn*, 1996; *Richardson and Cane*, 2004). Such a low ratio may be caused by crossing of the flank of the ropes (*Gopalswamy,* 2016; *Zhang et al.*, 2013), or it does reflect some progress destructing these rope structures. Previous studies have shown that CMEs will interact with ambient solar wind or adjacent CMEs as they propagate in the interplanetary space (*Farrugia and Berdichevsky*, 2004; *Dasso et al., 2007; Feng et al.*, 2009, 2011; *Gopalswamy et al.*, 2001, 2002; *Lavraud et al.*, 2014; *Liu et al.*, 2012, 2014a, 2014b; *Lugaz et al.*, 2012; *Manchester et al.*, 2014; *Mao et al.*, 2017; *Odstrcil et al.*, 2003; *Ruffenach et al.*, 2012, 2015; *Temmer et al.*, 2012). The interactions can alter the magnetic topology of CMEs and may result in great geomagnetic storms (*Liu et al.*, 2014b). However, due to expansion or the inhomogeneity in CMEs, the substructures (such as magnetic flux ropes) within CMEs are very likely to interact with each other. Therefore, understanding the interactions of these substructures is necessary for our understanding of the evolution of CMEs.

In this letter, we report an ICME consisting of multiple magnetic flux ropes**.** These ropes were merging with each other. The observations indicate that formation of large-scale rope structures can take place inside ICMEs as they are propagating in the interplanetary space. Our observations will provide improved understanding of the interplanetary evolution of CMEs.

## 2. Data

The data used in this paper are obtained from several instruments onboard Wind. The magnetic field data with time resolutions of 92s and 3s are taken from the Fluxgate Magnetometer experiment (*Lepping et al.,* 1995). The plasma data with time resolutions of 92s and 3s are respectively from The Solar Wind Experiment (*Ogilvie*

*et al.*, 1995) and the 3DP instrument (*Lin et al.*, 1995). If not specified, the GSE coordinate system (the Geocentric Solar Ecliptic coordinate system in which the $x$-axis directs from the Earth to the Sun, the $z$-axis points north, perpendicular to the ecliptic plane, the $y$-axis completes the right-handed coordinate system) is used in this paper.

## 3. Observations

Figure 1 shows observations of the solar wind made by WIND on March 25 and 26, 1998. From top to bottom, the magnitude and three components of the magnetic field, density, and temperature of protons and the three components of proton velocity and the proton plasma beta values are present. From ~12:00 March 25 to ~09:50 March 26 (i.e., shaded region), the WIND spacecraft encountered a region with relatively strong and smooth magnetic field and low proton temperature (Figures 1a–1d and 1f). During roughly the same interval, the plasma beta values were low (mostly below 0.1, Figure 1j), and the superthermal electrons were evidently bidirectional (data not shown here). Therefore, the shaded interval corresponded to an ICME (*Burlaga and King*, 1979; *Gosling and McComas*, 1987; *Richardson and Cane*, 1995).

From ~13:30 March 25 to 09:30 March 26, WIND observed two bipolar variations in $B_z$ (denoted by two red bars in the bottom of figure 1d). Within the bipolar $B_z$, the strength of $B_y$ and the total magnetic field reached their maxima (Figures 1a, 1c). A bipolar field with a core field in its center is the typical signature of the crossing of a magnetic flux rope. Therefore, WIND detected two flux ropes within the ICME. Applying G-S reconstruction method (*Hu and Sonnerup*, 2002) to the two red bar-denoted intervals, the axis of the two flux ropes were found in the direction of $\varphi = 260°, \theta = -10°$ for the earlier (i.e. FR1) and $\varphi = 96°, \theta = 19°$ for the latter, where $\varphi$ and $\theta$ are the longitude and latitude with respect to the ecliptic plane. The cross sections of the two flux ropes are shown in figures 2a and 2c.

Around 21:15 March 25, within the latter big flux rope (the one of longer duration), the spacecraft detected a local drop in $|B|$ and $|B_y|$. In the meantime, the density and the beta values reached a local maximum. Surveying the rope closely, the whole bipolar $B_z$ actually consisted of two successive bipolar variations (Figure 1d). Based on characteristics of the magnetic field and plasma, we conclude the latter big flux rope was formed by interaction of two small ropes denoted by FR2 and FR3 in figure 1d, respectively. The two small ropes are also reflected in the detailed version of the cross section near $y = 0$ (figure 2e). Between the FR1 and FR2, the spacecraft observed a simultaneous sharp reversal in $B_y$ and $B_z$ with a magnitude of ~15 nT, which corresponded to crossing of a current sheet. The details around the current sheet are shown in Figure 3.

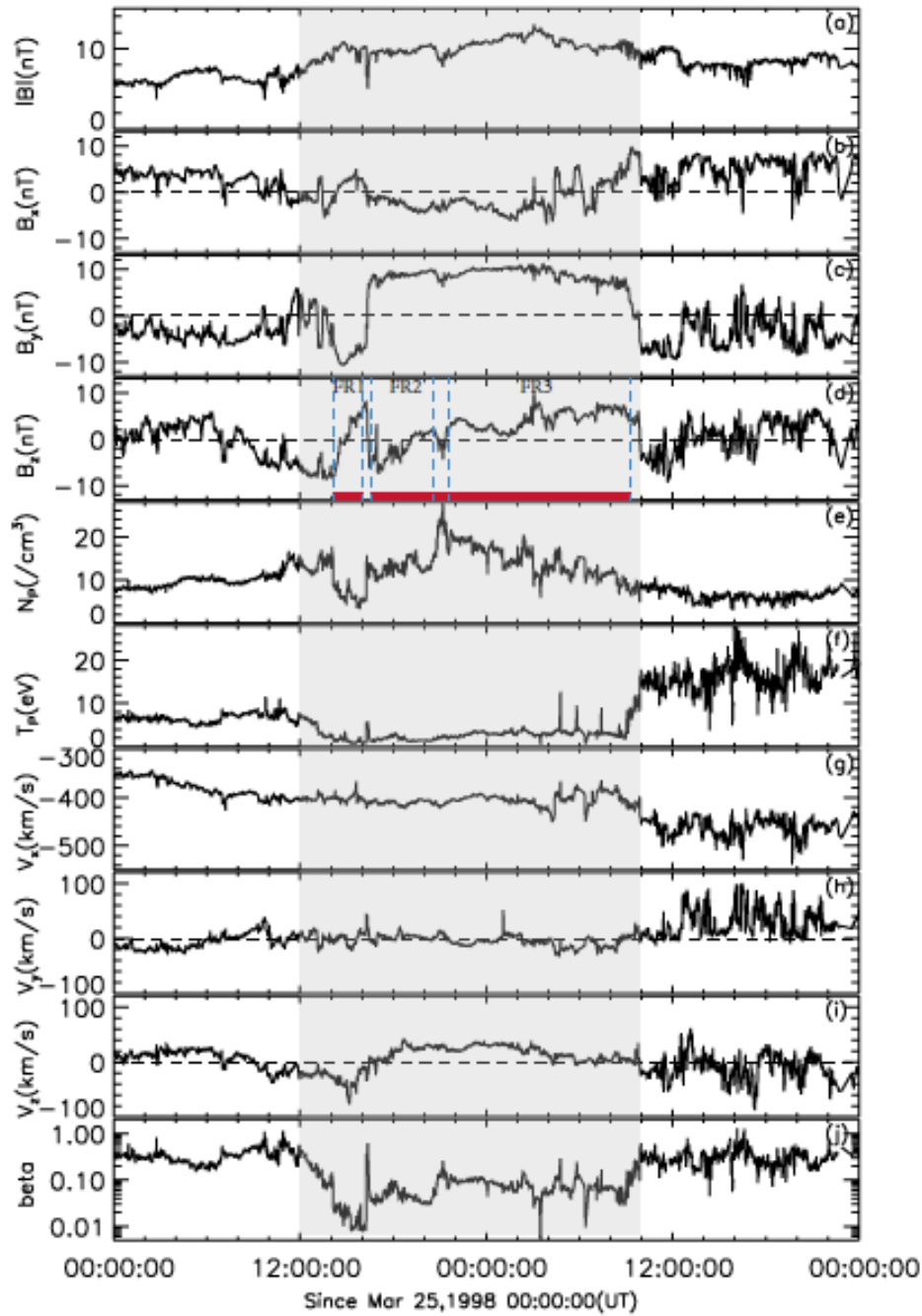

**Figure 1.** WIND's measurements from 00:00 UT on March 25 to 24:00 UT on March 26, 1998. (a–d) Magnitude and three components of the magnetic field. (e–f) Proton density and temperature. (g–i) Three components of plasma velocity. (j) The proton plasma beta values. Shaded region indicates the interval of ICME. The short red bar denotes the rope, FR1. The long red bar denotes the long-duration rope formed by the interaction of two flux ropes, FR2 and FR3, which are denoted by the two pairs blue dashed lines.

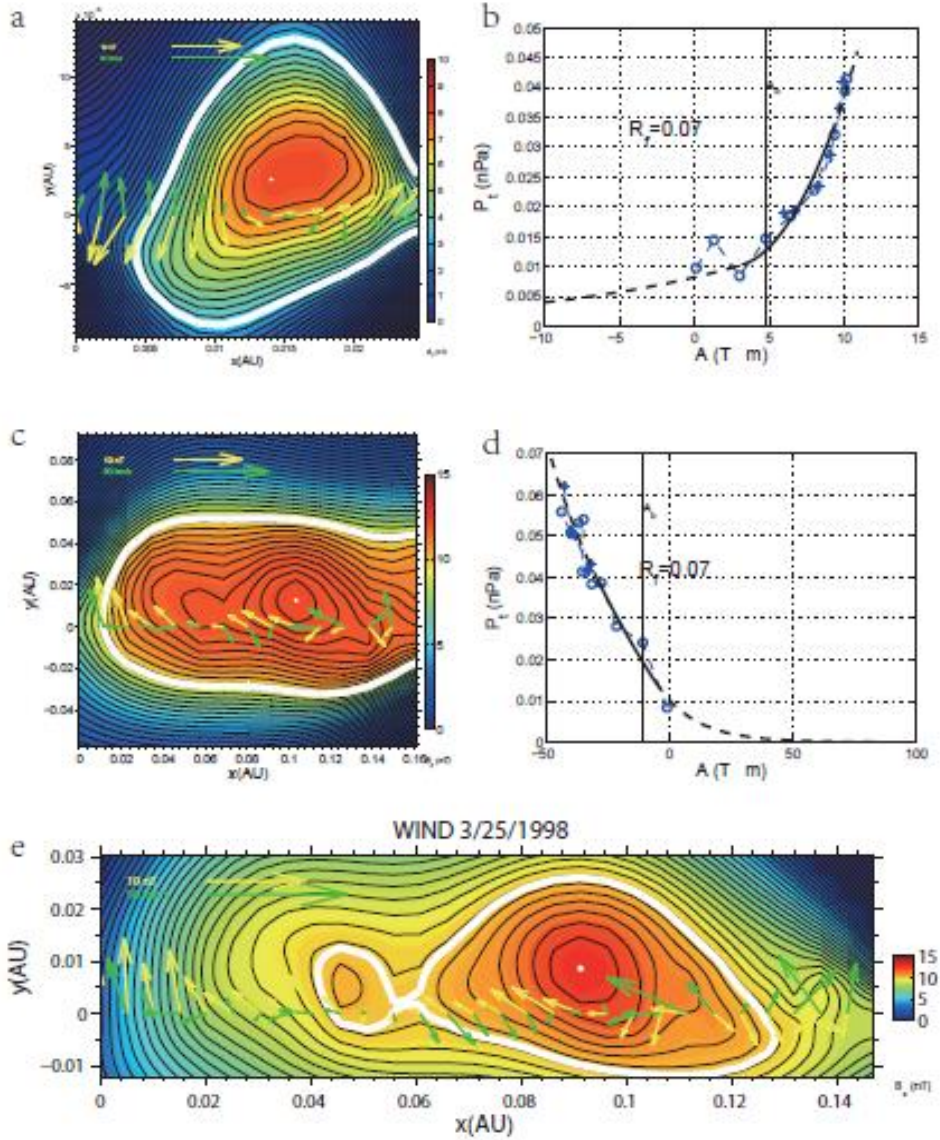

**Figure 2.** Grad–Shafranov reconstruction of the two intervals denoted by the two red bar in figure 1d with a-b for the short red bar (FR1) and c-e for the long red bar (FR2 and FR3). The left column shows the cross-sectional map of $A(x,y)$ (black contour lines) and the axial magnetic field $B_z$ (filled contours in color). Yellow arrows along $y=0$ denote measured transverse magnetic field vectors (scales given by the yellow arrow with a magnitude of 10 nT). White dot denotes the axial magnetic field maximum. The right column presents the measured transverse total pressure $P_t(x,0)$ (blue circles: inbound spacecraft path and stars: outbound spacecraft path) as a function of $A(x,0)$ and fitted $P_t(A)$ (thick black line). $R_f$ is the fitting residue value, which acts as an estimator of the reconstruction quality.

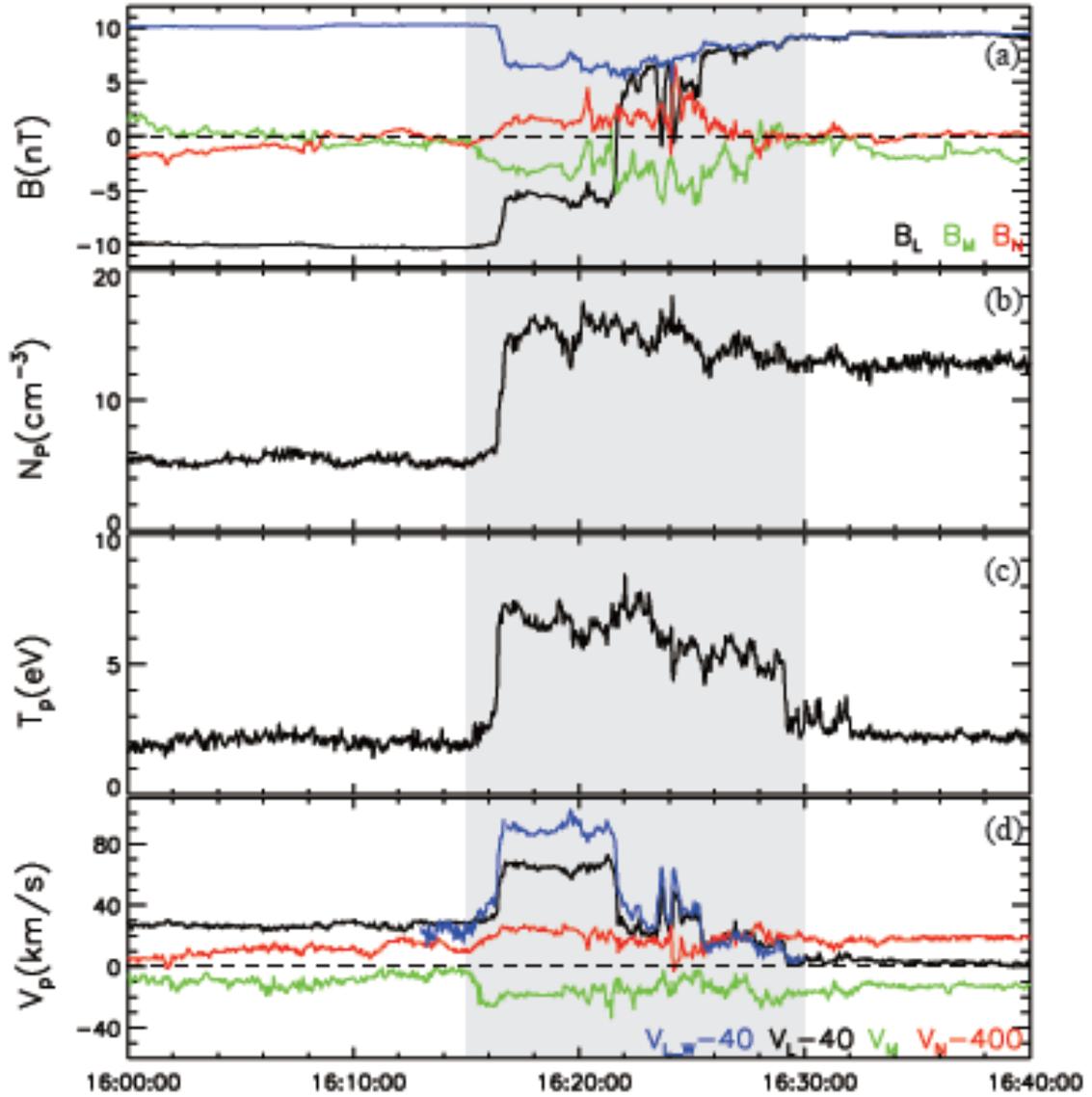

**Figure 3.** Expansion of the region between FR1 and FR2. (a) the magnetic field, (b-d) density and temperature, the plasma proton velocity. The shadow region denotes the reconnection current sheets. The magnetic field and the plasma proton velocity are pressed in the LMN coordinate system with $L = [-0.107, 0.703, -0.703]$, $M = [-0.0003, -0.707, -0.707]$, $N = [-0.994, -0.076, 0.076]$. In the LMN coordinate system, $N = (B_1 \times B_2)/|B_1 \times B_2|$ is assumed to be the current sheet normal, where $B_1$ and $B_2$ are the reference magnetic field selected respectively in the adjacent two side of the current sheet. $M = L` \times N$, where $L`$ is the maximum variance direction obtained from BMVA analysis. $L = M \times N$, is meant to be along the exhaust outflow direction.

In Figure 3, data with higher time resolution (3 s) are presented and the vectors are transformed to the *LMN* coordinate system which are obtained by the hybrid Minimum Variance analysis (*Gosling and Phan*, 2013; *Sonnerup and Cahill*, 1967). Corresponding to the current sheet between 16:15 and 16:30 (shaded region), the proton density and the temperature increased (Figures 3b–3c). The enhancement of the plasma velocity was mainly in the $L$ direction within the current sheet (Figure 3d, the black curve). The blue curve shown in Figure 3d present the tangential velocity derived from Walén relation (*Paschmann et al.*, 1896). The observed $V_L$ agreed well with that predicted by the Walén relation during the crossing of the current sheet. The magnitude of the observed $V_L$ only reached 70% of that predicted, which might derive from the isotropy assumption (which was used here) and the ignoring of heavy ions (*Paschmann et al.*, 1896). Therefore, the spacecraft detected an exhaust of magnetic reconnection located at the interfaces between FR1 and the latter big flux rope (*Gosling et al.*, 2005). Two adjacent flux ropes may merge with each other through magnetic reconnection and form a bigger one (*Odstrcil et al.*, 2003; Schmidt and Cargill, 2004). The identification of reconnection current sheet (i.e. the one around 16:20 March 25) between FR1 and the latter big flux rope indicated that the two flux ropes were merging through magnetic reconnection.

## 4. Discussion and Summary

In summary, we reported observations of an ICME within which two flux ropes with different durations (i.e. ~2 hours and ~17 hours, denoted by the two red bars in the bottom of figure 1d) were merging with each other through magnetic reconnection. The axis of the two flux ropes denoted by the two red bars in figure 1d were almost in the opposite direction. Therefore, the axial field ($B_y$) was also reconnected during their coalescence. Such a coalescence will form a bigger rope but with a weak axial field in the border which is often observed in Magnetic Clouds (MCs). The latter big rope (the long-duration one) was found to be formed by interaction of two smaller flux ropes (i.e. FR2 and FR3). A boundary region was formed and observed between FR2 and FR3, and it has similar properties (depression of the magnetic field strength, high proton density and plasma beta) of MC boundary layers formed through the interaction between MCs and the ambient medium (*Wei et al*. 2003).

Examining the CDAW LASCO CME catalog (*Yashiro et al.*, 2004) and taking factors of time window, central position angle and angular width into account, we found that the reported ICME may be a counterpart of the CME appeared in the FOV of SOHO/LASCO C2 (*Brueckner et al.*, 1995) at 07:31 UT on 21 Mar 1998. According to GOES X-Rays observations, we find only one flare took place at about 05:40 UT from the same source region. Thus, like the multi-flux-rope configuration CME reported by Awasthi et al. (2018), this CME may have multi-flux-rope configuration near the Sun. The observations of merging magnetic flux ropes demonstrate how the sub-structures of CMEs evolve. The multi-flux-rope ICMEs have been frequently reported in the early literature (*Fainberg et al.*, 1996; *Osherovich et al.*, 1997, 1998, 1999; *Ruzmaikin et al.*, 1997). Therefore, the interactions of sub-structures with CMEs are an important phenomenon in the evolution of CME. The observations provide strong support to the simulation results that two interacting ropes can gradually coalesce into one rope through magnetic reconnection when they propagate from the Sun to interplanetary medium (*Odstrcil et al.*, 2003; *Schmidt and Cargill*, 2004). In addition, the interactions of successive CMEs are often observed when they propagate in the heliosphere and many CMEs have magnetic flux rope structure. So, the reported rope–rope merger may be helpful for us to understand the CME-CME interaction and the evolution of CMEs. We speculated that the reported rope-rope interactions may also exist between successive rope-like CMEs and are important for the space weather forecasting.

**Acknowledgments**

We acknowledge supports from NSFC under grant Nos. 41674170. This work is also supported in part by the Plan for Scientific Innovation Talent of Henan Province (174100510019). We thank NASA/GSFC for the use of data from the Wind. These data can obtain freely from the Coordinated Data Analysis Web (http://cdaweb.gsfc.nasa.gov/cdaweb/istp_public/). We also acknowledge the use of data from SoHO/LASCO.

# References

Awasthi, A. K., Liu, R., Wang, H., Wang, Y., & Shen, C. (2018). Pre-eruptive Magnetic Reconnection within a Multi-flux-rope System in the Solar Corona. *Astrophys. J., 857*, 124. doi: 10.3847/1538-4357/aab7fb

Bothmer, V., Schwenn, R. (1996). Signatures of fast CMEs in interplanetary space, *Adv. Space Sci., 17*, 319– 322. doi: 10.1016/0273-1177(95)00593-4

Brueckner, G. E., et al. (1995). The Large Angle Spectroscopic Coronagraph (LASCO), *Solar Physics, 162*, 357

Brueckner, G. E. et al. (1998). Geomagnetic storms caused by coronal mass ejections (CMEs): March 1996 through June 1997. *Geophys. Res. Lett., 25*, 3019–3022. doi:10.1029/98GL00704

Burlaga, L. F., & King, J. H. (1979). Intense interplanetary magnetic fields observed by geocentric spacecraft during 1963-1975. *J. Geophys. Res., 84(A11)*, 6633-6640. doi: 10.1029/JA084iA11p06633

Canfield, R. C., Hudson, H. S., & Mckenzie, D. E. (1999). Sigmoidal morphology and eruptive solar activity. *Geophys. Res. Lett., 26(6)*, 627–630

Dasso, S., Nakwacki, M. S., Démoulin, P., & Mandrini, C. H. (2007). Progressive Transformation of a Flux Rope to an ICME. Comparative Analysis Using the Direct and Fitted Expansion Methods. *Solar Phys., 244*, 115-137. doi: 10.1007/s11207-007-9034-2

Fainberg, J. et al. (1996). Ulysseso bservationso f electron and proton components in a magnetic cloud and related wave activity, in *Proc. Solar Wind 8 Conference, Dana Point*, edited by D. Winterhalter, J.T. Gosling, S.R. Habbal, W.S. Kurth and M. Neugebauer, pp. 554-557

Farrugia, C., & Berdichevsky, D. (2004). Evolutionary signatures in complex ejecta and their driven shocks. Ann. Geophys., 22, 3679–3698. doi: 10.5194/angeo-22-3679-2004

Feng, H. Q., Wu, D. J. (2009). Observations of a small interplanetary magnetic flux

rope associated with a magnetic reconnection. *Astrophys. J., 705*, 1387. doi: 10.1088/0004-637X/705/2/1385

Feng, H. Q., Wu, D. J., Wang, J. M., Chao, J. K. (2011). Magnetic reconnection exhausts at the boundaries of small interplanetary magnetic flux ropes. *Astron. Astrophys., 527*, A67. doi:10.1051/0004-6361/201014473

Gopalswamy, N., Yashiro, S., Kaiser, M. L., Howard, R. A., and Bougeret, J. L. (2001). Radio signatures of coronal mass ejection interaction: Coronal mass ejection cannibalism?. *Astrophy. J., 548*, L91– L94. doi: 10.1086/318939

Gopalsamy, N., Yashiro, S., Kaiser, M. L., Howard, R. A., and Bougeret, J. L. (2002). Interplanetary radio emission due to interaction between two coronal mass ejections. *Geophys. Res. Lett., 29*, 1265–1268. doi: 10.1029/2001GL013606

Gopalswamy, N. (2016). History and development of coronal mass ejections as a key player in solar terrestrial Relationship. *Geosci. Lett., 3*, 8, doi:10.1186/s40562-016-0039-2

Gosling, J. T., & Mccomas, D. J. (1987). Field line draping about fast coronal mass ejecta: a source of strong out of the ecliptic interplanetary magnetic fields. *Geophys. Res. Lett., 14(4)*, 355–358. doi: 10.1029/GL014i004p00355

Gosling, J. T., R. M. Skoug, D. J. McComas, and C. W. Smith. (2005). Direct evidence for magnetic reconnection in the solar wind near 1 AU. *J. Geophys. Res., 110*, A01107, doi:10.1029/2004JA010809

Gosling, J. T., and T. D. Phan (2013). Magnetic reconnection in the solar wind at current sheets associated with extremely small field shear angles, *Astrophys. J. Lett., 763(2)*, L39, doi:10.1088/2041-8205/763/2/L39

Hu, Q., & Sonnerup, B. U. Ö. (2002). Reconstruction of magnetic clouds in the solar wind: orientation and configuration. *J. Geophys. Res., 107(A7)*, 10-15. doi: 10.1029/2001JA000293

Lavraud, B., Ruffenach, A., Rouillard, A. P., Kajdic, P., Manchester, W. B., Lugaz, N. (2014). Geo-effectiveness and radial dependence of magnetic cloud erosion by magnetic reconnection. *J. Geophys. Res., 119*, 26–35. doi: 10.1002/2013JA019154

Lepping, R. P., et al. (1995). The Wind Magnetic Field Investigation, edited by

Russell, C. T.. *Space Sci. Rev.*, *71*, 207–229. doi: 10.1007/BF00751330

Lin, R. P., et al. (1995). A three-dimensional plasma and energetic particle investigation for the wind spacecraft. *Space Sci. Rev., 71*, 125–153. doi:10.1007/BF00751328

Liu, R., Chang, L., Wang, S., Na, D., & Wang, H. (2010). Sigmoid-to-flux-rope transition leading to a loop-like coronal mass ejection. *Astrophys. J. Lett., 725(1)*, L84-L90

Liu, Y. D. et al. (2012). Interactions between coronal mass ejections viewed in coordinated imaging and in situ observations. *Astrophys. J. Lett., 746*, L15. doi: 10.1088/2041-8205/746/2/L15

Liu, Y. D., Yang, Z., Wang, R., Luhmann, J. G., Richardson, J. D., Lugaz, N. (2014a). Sun-to-Earth characteristics of two coronal mass ejections interacting near 1 au: formation of a complex ejecta and generation of a two-step geomagnetic storm. *Astrophys. J., 793*, L41–L46. doi: 10.1088/2041-8205/793/2/L41

Liu, Y. D., Luhmann, J. G., Kajdivc, P., et al. (2014b). Observations of an extreme storm in interplanetary space caused by successive coronal mass ejections. *Nature Communications, 5*, 3481. doi: 10.1038/ncomms4481

Lugaz, N. et al. (2012). The deflection of the two interacting coronal mass ejections of 2010 May 23-24 as revealed by combined in situ measurements and heliospheric imaging. *Astrophys. J., 759*, 68. doi:10.1088/0004-637X/759/1/68

Manchester, W. B., Kozyra, J. U., Lepri, S. T., Lavraud, B. (2014). Simulation of magnetic cloud erosion during propagation. *J. Geophys. Res., 119*, 5449–5464. doi: 10.1002/2014JA019882

Mao, S., He, J., Zhang, L., Yang, L., & Wang, L. (2017). Numerical Study of Erosion, Heating, and Acceleration of the Magnetic Cloud as Impacted by Fast Shock. *Astrophys. J. 842*, 109. doi: 10.3847/1538-4357/aa70e0

Odstrcil, D., Vandas, M., Pizzo, V. J., and MacNeice, P. (2003). Numerical simulation of interacting magnetic fluc ropes. *SOLARWIND 10*, edited by: Velli, M., Bruno, R., and Malara, F., 699–702. doi: 10.1063/1.1618690

Ogilvie, K. W., Chornay, D. J., Fitzenreiter, R. J., et al. (1995). SWE, A comprehensive plasma instrument for on the Wind spacecraft, edited by: Russell, C. T.. *Space Sci. Rev., 71*, 55–77. doi: 10.1007/BF00751326

Osherovich, V.A., et al. (1997). Self-similar evolution of Interplanetary magnetic clouds and Ulysses measurements of the polytropic index inside the cloud, in *The 31st ESLAB Symposium on Correlated Phenomena at the Sun*, in the Heliospherea nd in Geospacep, p. 171.

Osherovich, V.A. et al. (1998). Measurements of polytropic index in the January 10-11, 1997 magnetic cloud observed by Wind, *Geophys. Res. Lett. 25(15)*, 3003-3006

Osherovich, V. A., J. Fainberg, and R. G. Stone (1999). Multi‐tube model for interplanetary magnetic clouds, *Geophys. Res. Lett., 26*, 401–404, doi:10.1029/1998GL900306

Paschmann, G., Papamastorakis, I., Baumjohann, W., Sckopke, N., Carlson, C. W., & Sonnerup, B. U. Ö., et al. (1986). The magnetopause for large magnetic shear: AMPTE/IRM observations. *J. Geophys. Res., 91(A10)*, 11099-11115. doi: 10.1029/JA091iA10p11099

Richardson, I. G. & Cane, H. V. (1995). Regions of abnormally low proton temperature in the solar wind (1965–1991) and their association with ejecta. *J. Geophys. Res., 100(A12)*, 23397-23412. Doi: 10.1029/95JA02684

Richardson, I. G. & Cane, H. V. (2004). The fraction of interplanetary coronal mass ejections that are magnetic clouds: Evidence for a solar cycle variation. *Geophys. Res. Lett., 31*, L18804. doi: 10.1029/2004GL020958

Ruffenach, A., et al. (2012). Multispacecraft observation of magnetic cloud erosion by magnetic reconnection during propagation. *J. Geophys. Res., 117*, A09101. doi: 10.1029/2012JA017624

Ruffenach, A. et al. (2015). Statistical study of magnetic cloud erosion by magnetic reconnection. *J. Geophys. Res., 120*, 43-60. doi: 10.1002/2014JA020628

Rust, D. M., & Kumar, A. (2009). Evidence for helically kinked magnetic flux ropes in solar eruptions. *Astrophysical Journal, 464(2)*, L199

Ruzmaikin, A., J. Feynman and E.J. Smith (1997). Turbulence in mass ejections. *J.*

*Geophys. Res., 102*, 19753-19759.

Schmidt, J. & Cargill, P. (2004). A numerical study of two interacting coronal mass ejections. *Ann. Geophys., 22*, 2245–2254. doi: 10.5194/angeo-22-2245-2004

Sonnerup, B. U., and L. J. Cahill (1968). Explorer 12 observations of magnetopause current layer, *J. Geophys. Res., 73*, 1757–1770, doi:10.1029/JA073i005p01757

Temmer, M. et al. (2012). Characteristics of kinematics of a coronal mass ejection during the 2010 August 1 CME-CME interaction event. *Astrophys. J., 749*, 57. doi: 10.1088/0004-637X/749/1/57

Tsurutani, B. T., Gonzalez, W. D., Tang, F., Akasofu, S. I. and Smith, E. J. (1988). Origin of interplanetary southward magnetic fields responsible for major magnetic storms near solar maximum (1978–1979). *J. Geophys. Res., 93*, 8519–8531. doi:10.1029/JA093iA08p08519

Wei, F., Liu, R., Fan, Q., and Feng, X. (2003). Identification of the magnetic cloud boundary layers, *J. Geophys. Res., 108(A6)*, 1263, doi:10.1029/2002JA009511

Yashiro, S., N. Gopalswamy, G. Michalek, O. C. St. Cyr, S. P. Plunkett, N. B. Rich, and R. A. Howard (2004). A catalog of white light coronal mass ejections observed by the SOHO spacecraft, *J. Geophys. Res., 109*, A07105, doi: 10.1029/2003JA010282

Zhang, J., Cheng, X., & Ding, M. (2012). Observation of an evolving magnetic flux rope before and during a solar eruption. *Nature Communications, 3(1)*. doi:10.1038/ncomms1753